\documentclass[12pt]{article}
\usepackage{graphicx}

\def\eqref#1{(\ref{#1})}
\renewcommand{\thebibliography}[1]
{\vskip 8mm\centerline{\normalfont\normalsize\bfseries References}
\vskip 5mm\list
{\arabic{enumi}.}{\small
\setlength{\itemsep}{3mm}
\setlength{\parsep}{0mm}   
\settowidth\labelwidth{[#1]}\leftmargini\labelwidth
\advance\leftmarginii\labelsep
\usecounter{enumi}}}

\def\fnote#1#2{\begingroup\def\thefootnote{#1}\footnote{#2}\addtocounter
{footnote}{-1}\endgroup}
\begin{document}

\hfill{UTTG-01-07}

\vspace{36pt}

\begin{center}
{\large {\bf {Tensor Microwave Background Fluctuations for Large Multipole Order}}}

\vspace{36pt}
Raphael Flauger\fnote{*}{Electronic address:
flauger@physics.utexas.edu} and Steven Weinberg\fnote{**}{Electronic address:
weinberg@physics.utexas.edu}\\
{\em Theory Group, Department of Physics, University of
Texas\\
Austin, TX, 78712}

\vspace{30pt}

\noindent
{\bf Abstract}
\end{center}
\noindent
We present approximate formulas for the tensor BB, EE, TT, and TE multipole coefficients for large multipole order $\ell$.  The error in using the approximate formula for the BB multipole coefficients is less than cosmic variance for $\ell>10$.  These approximate formulas make  various qualitative properties of the calculated multipole coefficients transparent: specifically, they show that, whatever values are chosen for cosmological parameters, the tensor EE multipole coefficients   will always be larger than the BB coefficients for all $\ell>15$, and that these coefficients will approach each other for $\ell\ll 100$.  These approximations  also make clear how these multipole coefficients depend on cosmological parameters.
 \vfill

\pagebreak

\def\BM#1{\mbox{\boldmath{$#1$}}}

\begin{center}
{\bf I. Introduction}
\end{center}

Tensor fluctuations are a prime target for future observations of the cosmic microwave background, because if detected they can provide a conclusive verification of the theory of inflation and a unique tool for exploring the details of this theory.  The contribution of these fluctuations to the correlations of temperature and polarization correlations is well known.  They have the multipole coefficients:\footnote{These formulas are equivalent to those of Zaldarriaga and Seljak \cite{ZS}.  Their gravitational wave amplitude $h$ and power spectral function $P_h(k)$ are related to our gravitational wave amplitude ${\cal D}_q(t)$ by $h\sqrt{P_h}={\cal D}/2$.  In consequence, their  function $\Psi\sqrt{P_h}$ is $1/4$ times our source function $\Psi$.}  
\begin{eqnarray}
&&C^T_{EE,\ell}= \pi^2T_0^2\int_0^\infty q^2\,dq \nonumber\\ &&~~~~
\times\left|\int_{t_1}^{t_0}dt\,P(t)\,\Psi(q,t)\,\left\{\left[12+8\rho\frac{\partial}{\partial\rho}-\rho^2+\rho^2\frac{\partial^2}{\partial\rho^2}\right]
\frac{j_\ell(\rho)}{\rho^2}\right\}_{\rho=q\,r(t)}\right|^2\;,\label{eq:eeex}\\
&&C^T_{BB,\ell}= \pi^2T_0^2\int_0^\infty q^2\,dq \nonumber\\ &&~~~~
\times\left|\int_{t_1}^{t_0}dt\,P(t)\,\Psi(q,t)\,\left\{\left[ 8\rho+ 2\rho^2\frac{\partial}{\partial\rho}\right]
\frac{j_\ell(\rho)}{\rho^2}\right\}_{\rho=q\,r(t)}\right|^2\;,\label{eq:bbex}\\
&&C^T_{TE,\ell}=-2\pi^2 T_0^2\sqrt{\frac{(\ell+2)!}{(\ell-2)!}}\int_0^\infty q^2\,dq\nonumber\\&&~~~~\times\int_{t_1}^{t_0}dt\,P(t)\Psi(q,t)\left\{\left[12+8\rho\frac{\partial}{\partial\rho}-\rho^2+\rho^2\frac{\partial^2}{\partial\rho^2}\right]
\frac{j_\ell(\rho)}{\rho^2}\right\}_{\rho=q\,r(t)}\nonumber\\&&~~~~\times
\int_{t_1}^{t_0}dt'\,d(q,t')
\left\{ \frac{j_\ell\Big(qr(t')\Big)}{q^2r^2(t')}\right\}\;,\label{eq:teex}\\
&&
C^T_{TT,\ell}=\frac{4\pi^2(\ell+2)!T_0^2}{(\ell-2)!}\int_0^\infty q^2\,dq\left|\int_{t_1}^{t_0} dt\;d(q,t)\,\frac{j_\ell\Big(qr(t)\Big)}{q^2 r^2(t)}\right|^2\;.\label{eq:ttex}
\end{eqnarray}
Our notation here is consistent with that of reference \cite{SW1}.  To be explicit: $T_0$ is the  microwave background temperature at the present time $t_0$; $P(t)=\omega_c(t)\exp[-\int_t^{t_0}\omega_c(t')\,dt']$ is the  probability distribution of last scattering, with $\omega_c(t)$ the photon collision frequency; $t_1$ is any time taken early enough before recombination so that any photon present at $t_1$ would have collided many times before the present; $r(t)=\int_t^{t_0} dt'/a(t')$ is the co-moving radial coordinate of a source from which light emitted at time $t$ would reach us at the origin at the present time $t_0$; and $\Psi(q,t)$ is the ``source function,'' which is customarily calculated from a hierarchy of equations for partial-wave amplitudes:
\begin{eqnarray}
&&\dot{\Delta}^{(T)}_{T,\ell}(q,t)+\frac{q}{a(2\ell+1)}\Big((\ell+1)\Delta^{(T)}_{T,\ell+1}(q,t)-\ell \Delta^{(T)}_{T,\ell-1}(q,t)\Big)
\nonumber\\&&~~~~=\Big(-2\dot{\cal D}_q(t)+\omega_c(t)\Psi(q,t)\Big)\,\delta_{\ell,0}-\omega_c(t)\Delta^{(T)}_{T,\ell}(q,t)\;,\label{eq:bt}\\[0.3cm]&&
\dot{\Delta}^{(T)}_{P,\ell}(q,t)+\frac{q}{a(2\ell+1)}\Big((\ell+1)\Delta^{(T)}_{P,\ell+1}(q,t)-\ell \Delta^{(T)}_{P,\ell-1}(q,t)\Big)
\nonumber\\&&~~~~=-\omega_c(t)\Psi(q,t)\,\delta_{\ell,0}-\omega_c(t)\Delta^{(T)}_{P,\ell}(q,t)\;,\label{eq:bp}
\end{eqnarray}
with
\begin{eqnarray}
&&\Psi(q,t)=\frac{1}{10}\Delta^{(T)}_{T,0}(q,t)+\frac{1}{7}\Delta^{(T)}_{T,2}(q,t)+
\frac{3}{70}\Delta^{(T)}_{T,4}(q,t)-\frac{3}{5}\Delta^{(T)}_{P,0}(q,t)\nonumber\\&&~~~~~~~+\frac{6}{7}\Delta^{(T)}_{P,2}(q,t)-\frac{3}{70}\Delta^{(T)}_{P,4}(q,t)\;.
\end{eqnarray}
Here ${\cal D}_q(t)$ is the gravitational wave amplitude (apart from terms that decay outside the horizon), defined by
\begin{equation}
\delta g_{ij}({\bf x},t)=\sum_\pm\int d^3q\;e^{i{\bf q}\cdot{\bf x}}\,\beta({\bf q},\pm 2)\,e_{ij}(\hat{q},\pm 2)\,{\cal D}_q(t)\;,
\end{equation}
with $\beta({\bf q},\pm 2)$ and $e_{ij}(\hat{q},\pm 2)$ the stochastic parameter and polarization tensor for helicity $\pm 2$, normalized so that 
\begin{equation}
\langle \beta({\bf q},\lambda)\,\beta^*({\bf q}',\lambda')\rangle=\delta_{\lambda\lambda'}\delta^3({\bf q}-{\bf q}')\;,
\end{equation}
and for $\hat{q}$ in the 3-direction
\begin{equation}
e_{11}(\hat{q},\pm 2)=-e_{22}(\hat{q},\pm 2)=1/\sqrt{2}\;,~~~
e_{12}(\hat{q},\pm 2)=e_{21}(\hat{q},\pm 2)=\pm i/\sqrt{2}\;.
\end{equation}
Finally, $d(q,t)$ is the quantity
\begin{equation}
d(q,t)\equiv \exp\left[-\int_t^{t_0}dt'\,\omega_c(t')\right]\,\left(\dot{\cal D}_q(t)-\frac{1}{2}\omega_c(t)\Psi(q,t)\right)\;.
\end{equation}

Aside from the treatment of the tensor mode as a first-order perturbation, and the assumption of purely elastic Thomson scattering, Eqs. \eqref{eq:eeex}--\eqref{eq:ttex} may be regarded as exact.  They serve as the basis of computer programs such as CMBfast and CAMB, that are used to compare observations of microwave background polarization and temperature fluctuations with models that predict values for the gravitational wave amplitude ${\cal D}_q(t)$.  But they are not very transparent.

\begin{figure}[h]
\begin{center}
\includegraphics[height=3.7in]{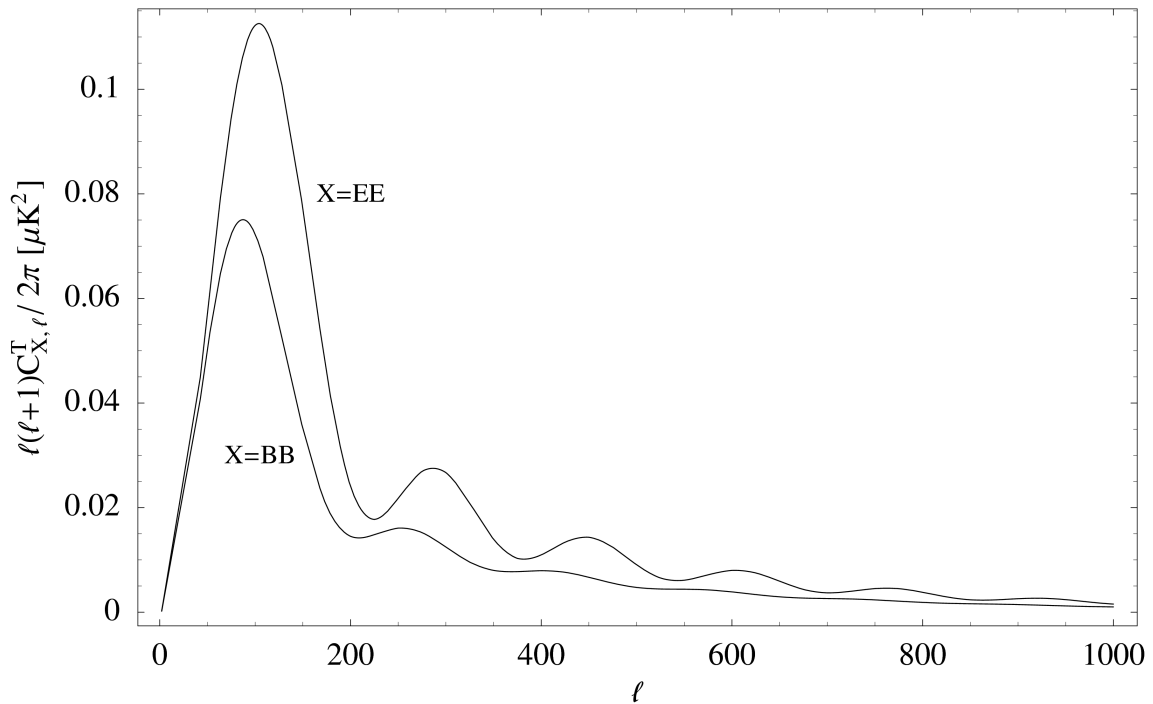}
\end{center}
\caption{Comparison of $C^T_{EE,\ell}$ and $C^T_{BB,\ell}$, in $(\mu {\rm K})^2$.}
\label{fig:eevsbb}
\end{figure}

	For one thing, as shown in Figure \ref{fig:eevsbb}, computer calculations using Eqs.~\eqref{eq:eeex} and \eqref{eq:bbex} yield results for  $C^T_{EE,\ell}$ and $C^T_{BB,\ell}$ that are of the same order of magnitude, and nearly equal for $\ell<100$, while $C^T_{EE,\ell}> C^T_{BB,\ell}$ for all $\ell>15$.  \\

       Of course computer calculations can only show this for specific choices of cosmological parameters.  (The cosmological parameters used in Figure 1 are described below.)  It would be impossible to conclude  just by inspection of Eqs.~\eqref{eq:eeex} and \eqref{eq:bbex} that these are general properties of the multipole coefficients, independent of the choice of cosmological parameters.  In this paper we present  successive approximations that make these properties apparent, and that, at the cost of only a small additional loss in accuracy, also clarify how the multipole coefficients depend on various cosmological parameters.  \\

\begin{center}
{\bf II. The Large-$\ell$ Approximation}
\end{center}

We can approximate Eqs.~\eqref{eq:eeex}--\eqref{eq:ttex} by much simpler and more transparent formulas, by using an asymptotic formula \cite{GR} for the spherical Bessel functions \footnote{After the preprint of this work was first circulated we learned that the same approximation is used by J. R. Pritchard and M. Kamionkowski, Ann. Phys.  {\bf 318}, 2 (2005) [astro-ph/0412581]. However,  after making this approximation they make further approximations that are quite different from ours, and that lead to a divergence in the integral over wave number, which must be dealt with by an arbitrary cut-off. The error introduced by their approximation is comparable to the one introduced by our last approximation given by Eqs. \eqref{eq:eeap3} and \eqref{eq:bbap3}. Another approximation that consists in averaging over the rapid oscillations in the square of the Bessel functions leading to results similar to our last approximation was proposed by M. Zaldarriaga and D. D. Harari, Phys. Rev. D {\bf 52}, 3276 (1995) [astro-ph/9504085]. An analytic expression for the contribution of the tensor modes to the temperature multipole coefficients approximately valid for $1\ll \ell < 50$ obtained using a similar average was given by A. A. Starobinsky, Sov. Astron. Lett. {\bf 11}, 133 (1985)}:
\begin{equation}
j_\ell(\rho)\rightarrow\left\{\begin{array}{cl} \frac{\cos b\,\cos\Big[\nu(\tan b- b)-\pi/4\Big]}{\nu\sqrt{\sin b}} &~~~~~~~ \rho>\nu \\ 0 &~~~~~~~ \rho<\nu \end{array}\right.\;,\label{eq:jlap}
\end{equation}
where $\nu\equiv \ell+1/2$, and $\cos b\equiv \nu/\rho$, with $0\leq b\leq \pi/2$.  This approximation is valid for $|\nu^2-\rho^2|\gg \nu^{4/3}$.  
Hence for $\ell\gg 1$, this formula can be used over most of the ranges of integration in Eqs. \eqref{eq:eeex}--\eqref{eq:ttex}.  Furthermore, for $\rho>\nu\gg 1 $ the phase $\nu(\tan b- b)$  in Eq. \eqref{eq:jlap} is a very rapidly increasing function of $\rho$, so the derivatives in Eqs. \eqref{eq:eeex}--\eqref{eq:teex} can be taken to act chiefly on this phase:
\begin{eqnarray}
&&\left[12+8\rho\frac{\partial}{\partial\rho}-\rho^2+\rho^2\frac{\partial^2}{\partial\rho^2}\right]
\frac{j_\ell(\rho)}{\rho^2}\rightarrow -j_\ell(\rho)+j''_\ell(\rho)\\\nonumber&&\hskip 4.52cm \rightarrow -\frac{(1+\sin^2 b)\,\cos b}{\nu\sqrt{\sin b}}
\cos\Big[\nu(\tan b- b)-\pi/4\Big]\\
&&
\left[ 8\rho+ 2\rho^2\frac{\partial}{\partial\rho}\right]
\frac{j_\ell(\rho)}{\rho^2}\rightarrow 2j'_\ell(\rho)\rightarrow -\frac{2\sqrt{\sin b}\cos b}{\nu}
\sin\Big[\nu(\tan b- b)-\pi/4\Big]
\end{eqnarray}
Then Eqs. \eqref{eq:eeex}--\eqref{eq:ttex} become, for $\nu=\ell+1/2\gg 1$,
\begin{eqnarray}
&&\ell(\ell+1)C^T_{EE,\ell}\rightarrow \pi^2T_0^2\int_0^\infty q^2\,dq \nonumber\\ &&~~~~
\times\left|\int_{r(t)>\nu/q}dt\,P(t)\,\Psi(q,t)\,\left\{\frac{(1+\sin^2 b)\,\cos b}{\sqrt{\sin b}}
\cos\Big[\nu(\tan b- b)-\pi/4\Big]\right\}_{\cos b=\nu/q\,r(t)}\right|^2\;,\label{eq:eeap1}\nonumber\\&&{}\\&&
\ell(\ell+1)C^T_{BB,\ell}\rightarrow \pi^2T_0^2\int_0^\infty q^2\,dq \nonumber\\ &&~~~~
\times\left|\int_{r(t)>\nu/q}dt\,P(t)\,\Psi(q,t)\,\left\{2\sqrt{\sin b}\cos b\,
\sin\Big[\nu(\tan b- b)-\pi/4\Big]\right\}_{\cos b=\nu/q\,r(t)}\right|^2\;,\label{eq:bbap1}\nonumber\\&&\\&&{}
\ell(\ell+1)C^T_{TE,\ell}\rightarrow-2\pi^2 T_0^2\int_0^\infty q^2\,dq\nonumber\\&&~~~~\times\int_{r(t)>\nu/q}dt\,P(t)\Psi(q,t)\left\{\frac{(1+\sin^2 b)\,\cos b}{\sqrt{\sin b}}
\cos\Big[\nu(\tan b- b)-\pi/4\Big]\right\}_{\cos b=\nu/q\,r(t)}\nonumber\\&&~~~~\times
\int_{r(t')>\nu/q}dt'\,d(q,t')\, \left\{\frac{\cos^3b}{\sqrt{\sin b}}
\cos\Big[\nu(\tan b- b)-\pi/4\Big]\right\}_{\cos b=\nu/q\,r(t')}\;,\label{eq:teap1}\\
&&\ell(\ell+1)C^T_{TT,\ell}\rightarrow 4\pi^2T_0^2\int_0^\infty q^2\,dq \nonumber\\&&~~~~\times\left|\int_{r(t)>\nu/q} dt\;d(q,t)\,\left\{\frac{\cos^3 b}{\sqrt{\sin b}}
\cos\Big[\nu(\tan b- b)-\pi/4\Big]\right\}_{\cos b=\nu/q\,r(t)}\right|^2\;.\label{eq:ttap1}
\end{eqnarray}\\

\begin{figure}[ht]
\begin{center}
\includegraphics[height=3.7in]{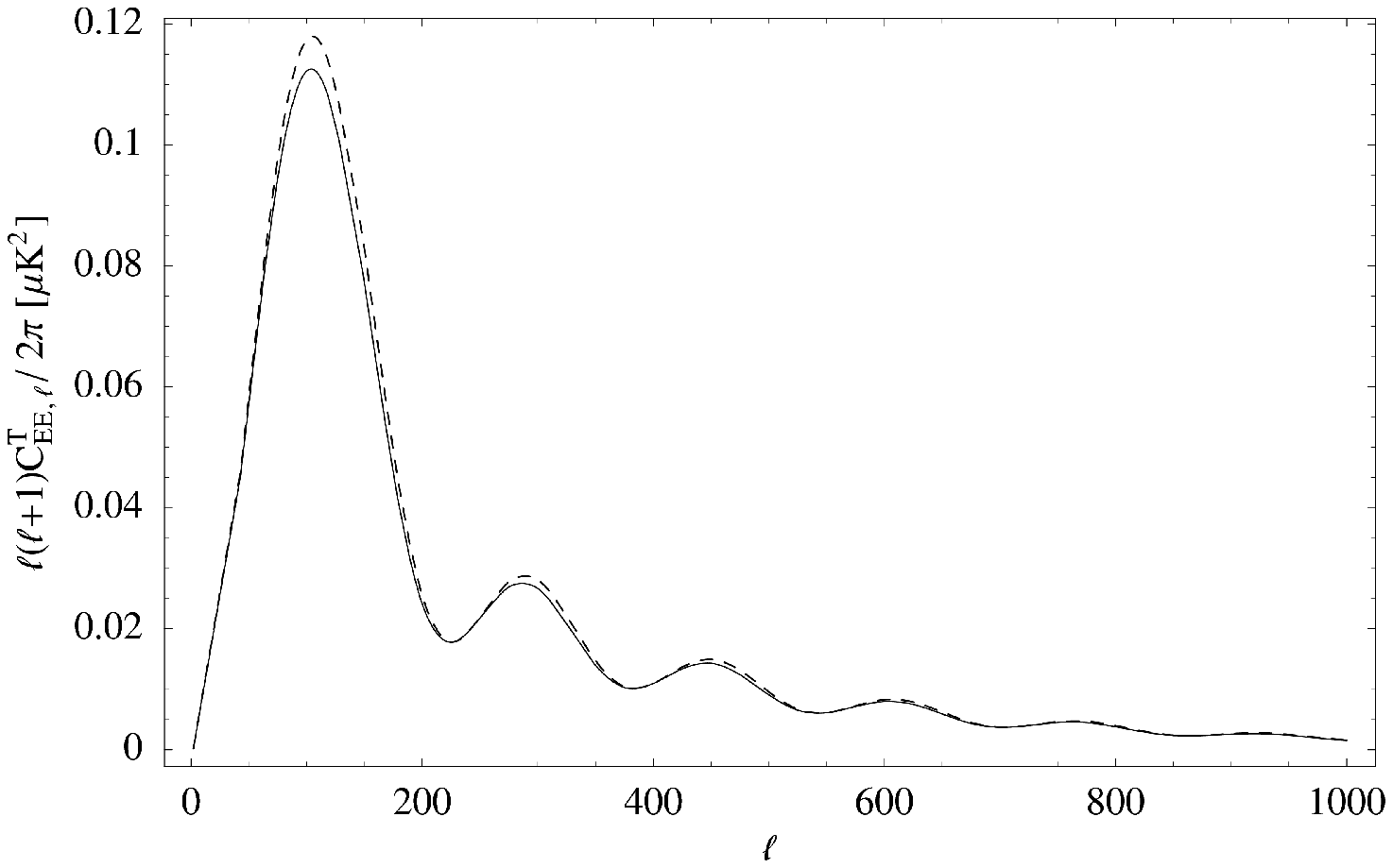}
\end{center}
\caption{Comparison of formulas for $C_{EE,\ell}^T$.  The solid line is the result of using the exact expression \eqref{eq:eeex}; the dashed line is the result of using the approximation \eqref{eq:eeap1}.  Figures 2--5 show the degree of accuracy of the large-$\ell$ approximation by itself, without further approximations.  In this and all other figures, all calculations are done using the cosmological parameters given in Section II, and the units of the vertical axis are square microKelvins.}
\label{fig:eeexvsap1}
\end{figure}

\begin{figure}[ht]
\begin{center}
\includegraphics[height=3.7in]{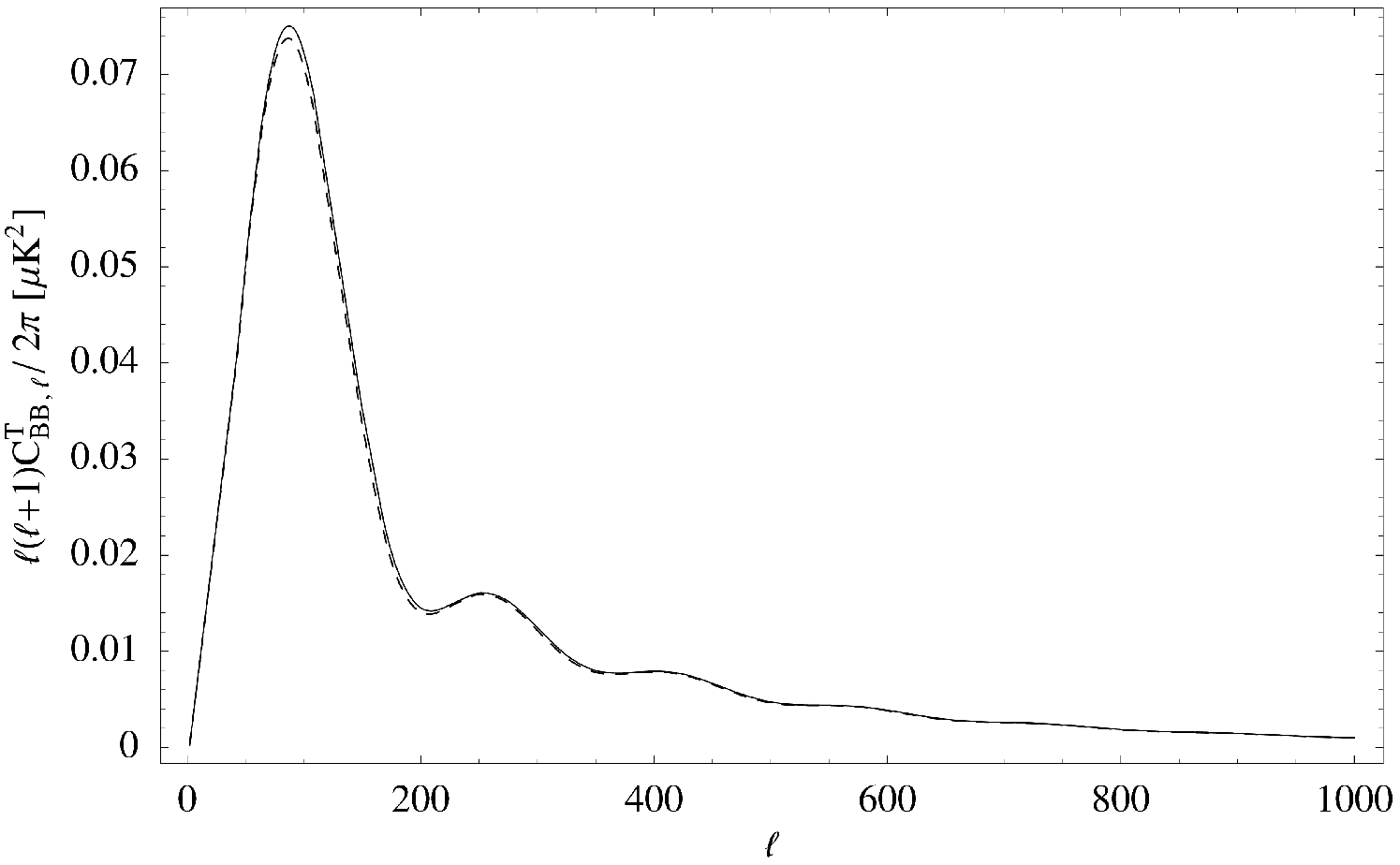}
\end{center}
\caption{Comparison of formulas for $C_{BB,\ell}^T$.  The solid line is the result of using the exact expression \eqref{eq:bbex}; the dashed line is the result of using the approximation \eqref{eq:bbap1}.}
\label{fig:bbexvsap1}
\end{figure}

\begin{figure}[ht]
\begin{center}
\includegraphics[height=3.7in]{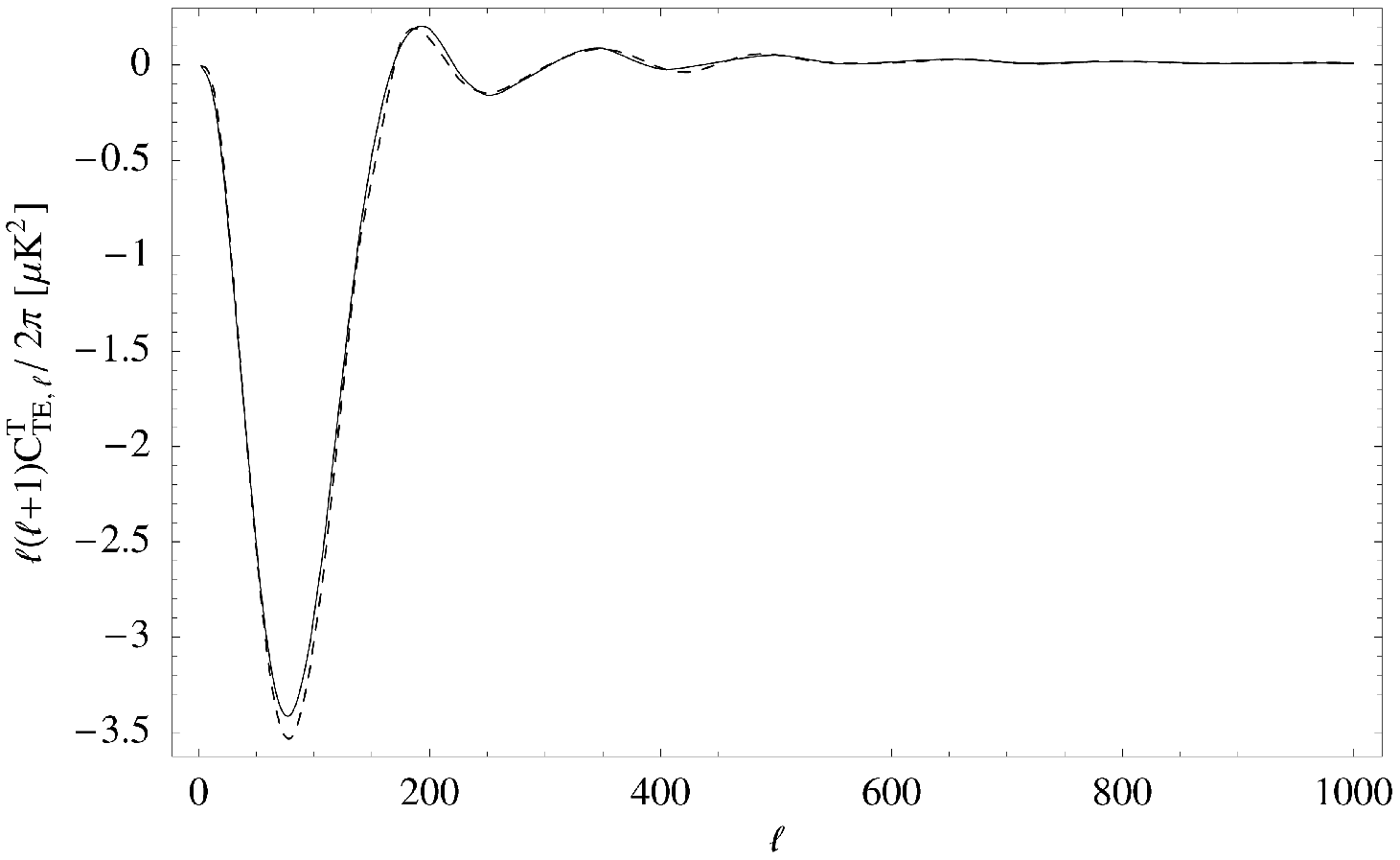}
\end{center}
\caption{Comparison of formulas for $C_{TE,\ell}^T$.  The solid line is the result of using the exact expression \eqref{eq:teex}; the dashed line is the result of using the approximation \eqref{eq:teap1}.}
\label{fig:teexvsap1}
\end{figure}

\begin{figure}[ht]
\begin{center}
\includegraphics[height=3.7in]{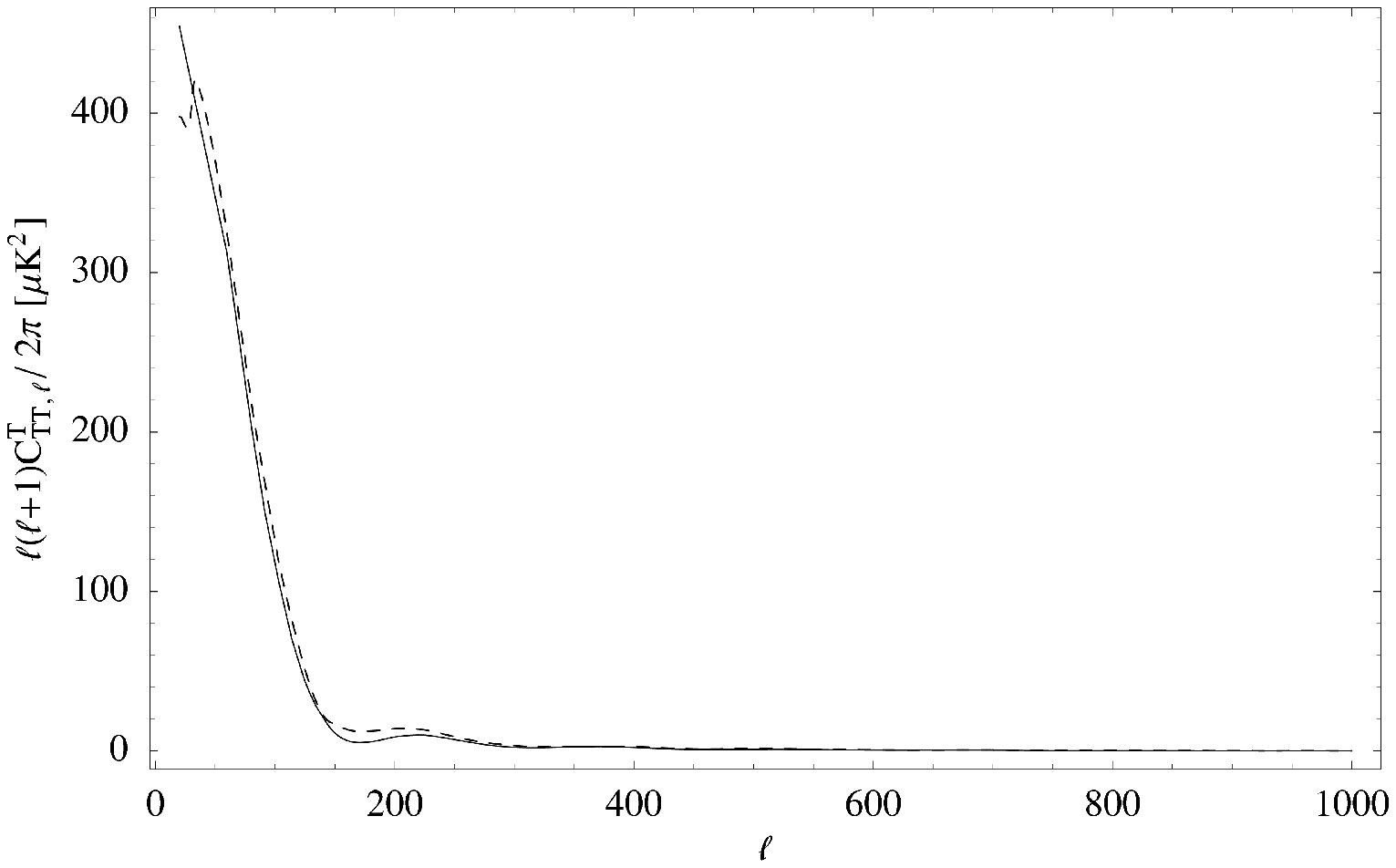}
\end{center}
\caption{Comparison of formulas for $C_{TT,\ell}^T$.  The solid line is the result of using the exact expression \eqref{eq:ttex}; the dashed line is the result of using the approximation \eqref{eq:ttap1}.}
\label{fig:ttexvsap1}
\end{figure}

In evaluating both the exact and  approximate expressions, instead of calculating the source function $\Psi(q,t)$ by truncating the Boltzmann hierarchy \eqref{eq:bt}, \eqref{eq:bp}, we use the  integral equation \cite{SW1,BGP}
\begin{eqnarray}
&&\Psi(q,t)=\frac{3}{2}\int_{t_1}^t dt' \exp\left[-\int_{t'}^{t}\omega_c(t'')\,dt''\right]
\nonumber\\&&~~~\times \Bigg[ -2\dot{\cal D}_q(t')K\left(q\int_{t'}^{t}\frac{dt''}{a(t'')}\right)
+\omega_c(t')F\left(q\int_{t'}^t\frac{dt''}{a(t'')}\right)\,\Psi(q,t')\Bigg]\;,\nonumber\\&&{}
\end{eqnarray}
where $K(v)$ and $F(v)$ are the functions
\begin{equation}
K(v)\equiv j_2(v)/v^2\;,~~~F(v)\equiv  j_0(v)-2j_1(v)/v+2j_2(v)/v^2\;.
\end{equation}
(This is not an approximation; in principle it should give the same results as the truncated Boltzmann hierarchy used by CMBfast and CAMB, aside from the supposedly small errors produced by the truncation.  In fact, our method gives results that differ by a few percent from both CMBfast and CAMB, but CMBfast and CAMB give results for both $C_{EE,\ell}^T$ and $C_{BB,\ell}^T$ that differ by similar amounts from each other, especially for large $\ell$.  At this point we are not able to tell which of the three methods is the most reliable.) 
The specific cosmological model chosen for this and all other numerical calculations in this paper is consistent with current observations:  We assume zero spatial curvature and constant vacuum energy, with density parameters for baryons, cold dark matter, and dark energy given by $\Omega_B=0.0432$, $\Omega_{CDM}=0.213$, $\Omega_\Lambda=0.743$.  We take the reduced Hubble constant as $h=0.72$ and the present microwave background temperature as $T_0=2.725$K.  In calculating the photon collsion frequency, we use the recfast recombination code \cite{REC}, with helium abundance $Y=0.24$.  The gravitational field amplitude outside the horizon is taken as 
$$|{\cal D}_q|^2=4.34\times 10^{-11}\,q^{-3}$$ corresponding to  $n_T=0$, $A=0.739$, and a tensor/scalar ratio $r=1$ in the notation of \cite{Spergel03}. (The values of the parameters used are the rounded maximum likelihood values from the full N-dimensional likelihood analysis and may differ slightly from the marginalized values quoted in \cite{Spergel03}). The gravitational wave amplitude ${\cal D}_q(t)$ is calculated including the damping due to neutrino anisotropic inertia, as in \cite{SW2}; the effects of photon anisotropic inertia are negligible.  Reionization is ignored.  To take into account a finite optical depth $\tau$ of the reionized plasma or a different value of $r$, for $\ell>10$ it is only necessary  to multiply the multipole coefficients given here by $r\exp(-2\tau)$.
The approximate results obtained in this way from Eqs. \eqref{eq:eeap1}--\eqref{eq:ttap1} are compared with the exact formulas \eqref{eq:eeex}--\eqref{eq:ttex} in Figures \ref{fig:eeexvsap1}--\ref{fig:ttexvsap1}.    

We can gain further simplicity and transparency in the formulas for the EE and BB multipole coefficients by using another approximation that actually leads to improved accuracy for $C_{BB,\ell}^T$.     The last-scattering probability distribution $P(t)$ is concentrated around a time $t_L$, corresponding to a redshift $z_L\simeq 1090$.    For any $q$ of the same order of magnitude as $\nu/r(t_L)$, the quantity $b\equiv \cos^{-1}\Big(\nu/qr(t)\Big)$ does not vary appreciably for $t$ within the range in which $P(t)$ is appreciable.   Hence we can set $r(t)$ equal to $r_L$ everywhere {\em except} in the phase $\nu(\tan b-b)$, which for $\nu\gg 1$ does vary over a wide range in this interval.  Furthermore, because $\nu(\tan b-b)$ varies over a wide range for $\nu\gg 1$, the difference between $\cos[\nu(\tan b-b)-\pi/4]$ and $\sin[\nu(\tan b-b)-\pi/4]$ is immaterial, and we can replace both with $\cos[\nu(\tan b-b)]$.  Making these replacements in Eqs. \eqref{eq:eeap1} and \eqref{eq:bbap1} gives
\begin{eqnarray}
&&\ell(\ell+1)C^T_{EE,\ell}\rightarrow \pi^2T_0^2\int_{\nu/r_L}^\infty q^2\,dq \,\{(1+\sin^2 b_L)^2\,\cos^2 b_L\}_{\cos b_L=\nu/qr_L}\nonumber\\ &&~~~~
\times\left|\int_{r(t)>\nu/q}dt\,P(t)\,\Psi(q,t)\,\left\{\frac{
\cos\Big[\nu(\tan b- b)\Big]}{\sqrt{\sin b}}\right\}_{\cos b=\nu/q\,r(t)}\right|^2\;,\label{eq:eeap2}\nonumber\\&&{}\\&&
\ell(\ell+1)C^T_{BB,\ell}\rightarrow \pi^2T_0^2\int_{\nu/r_L}^\infty q^2\,dq \,\{4\sin^2b_L\cos^2b_L\}_{\cos b_L=\nu/q\,r_L}\nonumber\\ &&~~~~
\times\left|\int_{r(t)>\nu/q}dt\,P(t)\,\Psi(q,t)\,\left\{\frac{
\cos\Big[\nu(\tan b- b)\Big]}{\sqrt{\sin b}}\right\}_{\cos b=\nu/q\,r(t)}\right|^2\;.\label{eq:bbap2}
\end{eqnarray}

(We have not set $b=b_L$ in the factors $1/\sqrt{\sin b}$ in both integrals over $t$, in order to avoid a divergence in the integration over $q$ at $q=\nu/r_L$.   This factor does not introduce a divergence in the integrals over time, because $dt\propto \sin b\,db$.)

\begin{figure}[ht]
\begin{center}
\includegraphics[height=3.7in]{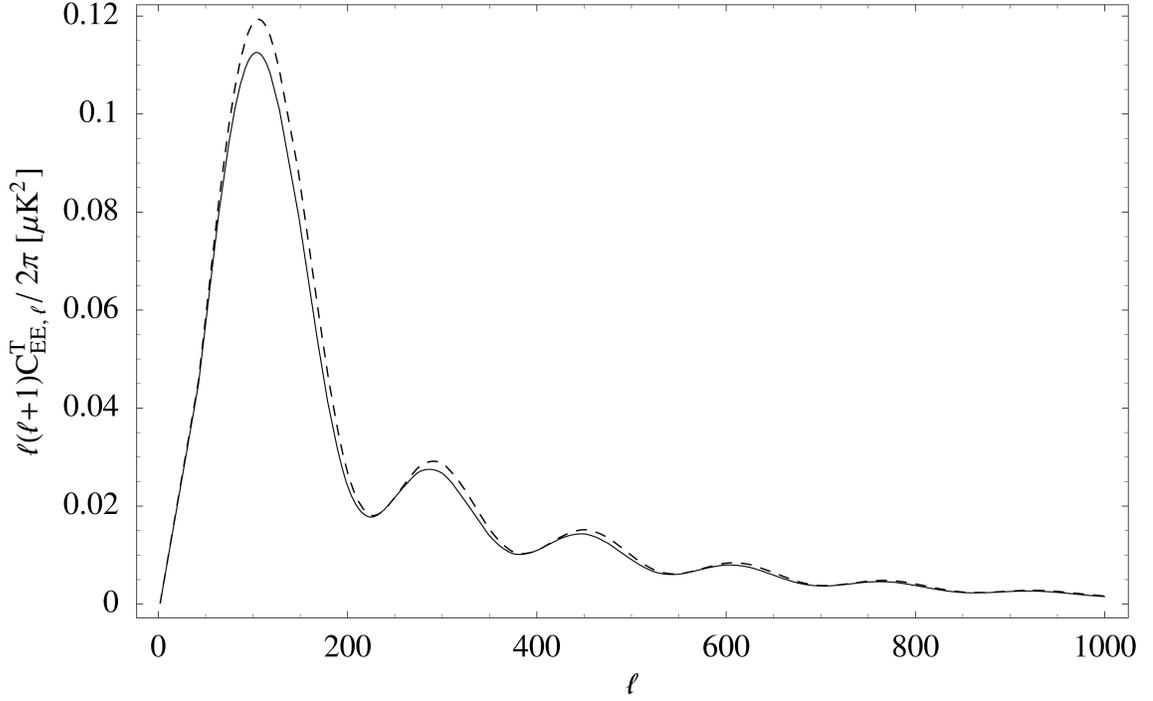}
\end{center}
\caption{Comparison of formulas for $C_{EE,\ell}^T$.  The solid line is the result of using the exact expression \eqref{eq:eeex}; the dashed line is the result of using the approximation \eqref{eq:eeap2}.  Figures 6 and 7 show the degree of accuracy of the combined approximations that we use to show analytically that $C_{EE,\ell}^T>C_{BB,\ell}^T$.}
\label{fig:eeexvsap2}
\end{figure}

\begin{figure}[ht]
\begin{center}
\includegraphics[height=3.7in]{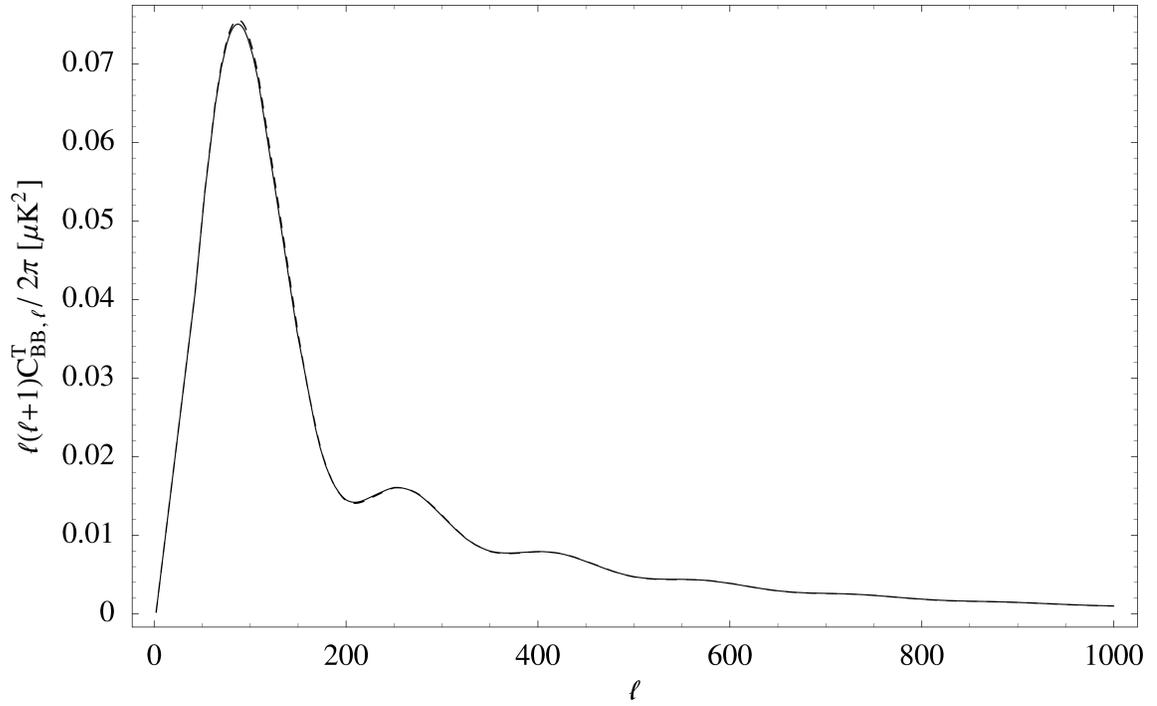}
\end{center}
\caption{Comparison of formulas for $C_{BB,\ell}^T$.  The solid line is the result of using the exact expression \eqref{eq:bbex}; the dashed line is the result of using the approximation \eqref{eq:bbap2}.  This is our best approximation for $C_{BB,\ell}^T$}
\label{fig:bbexvsap2}
\end{figure}

These approximate formulas are compared with results of the exact formulas \eqref{eq:eeex} and \eqref{eq:bbex} in Figures \ref{fig:eeexvsap2} and \ref{fig:bbexvsap2}.  The approximate result \eqref{eq:bbap2} for $C_{BB,\ell}^T$ agrees with the exact result \eqref{eq:bbex} to about 1\% for all $\ell>10$, which is better than cosmic variance.  The approximate result \eqref{eq:eeap2} for $C_{EE,\ell}^T$ is not quite as accurate; it agrees with the exact result \eqref{eq:eeex} to better than about 14\% for all $\ell>10$.  These approximations are evidently   accurate enough for us to draw qualitative conclusions about the EE and BB multipole coefficients.

  One immediate consequence is that, since $(1+\sin^2 b_L)^2\geq 4\sin^2 b_L$ for all real $b_L$, we expect that $C^T_{EE,\ell}\geq C^T_{BB,\ell}$ for all $\ell$ large enough  to justify our approximations.  Also, since $\Psi(q,t_L)$ falls off for wave lengths that come into the horizon before matter-radiation equality,  we expect that for relatively small $\ell$ (say, $\ell<100$) the integrals over $q$ are dominated by values for which $\cos b_L$ is small, so that $(1+\sin^2 b_L)^2\simeq 4\sin^2 b_L$, and hence $C^T_{EE,\ell}\simeq C^T_{BB,\ell}$ for such $\ell$.  As mentioned in Section I, and shown in Figure \ref{fig:eevsbb}, both properties are observed in the output of numerical calculations based on the accurate formulas \eqref{eq:eeex} and \eqref{eq:bbex}.  \\

\begin{center}
{\bf III. Parameter-Dependence of the EE and BB Correlations}
\end{center}

With one further approximation, we can find reasonably accurate formulas for $C^T_{EE,\ell}$ and $C^T_{BB,\ell}$ that reveal the way that these coefficients depend on various cosmological parameters.  We write the squared time integrals in Eqs. \eqref{eq:eeap2} and \eqref{eq:bbap2} as double integrals over times $t$ and $t'$, and write
\begin{eqnarray*}
&&\cos\Big[\nu(\tan b- b)\Big]\,\cos\Big[\nu(\tan b'- b')\Big]=\frac{1}{2}\Bigg[\cos\Big[\nu(\tan b- b)-\nu(\tan b'- b')\Big]\\&&~~~~~~~~~+\cos\Big[\nu(\tan b- b)+\nu(\tan b'- b')\Big]\Bigg]\;,
\end{eqnarray*}

where $\cos b=\nu/q\,r(t)$ and $\cos b'=\nu/q\,r(t')$.  For $\nu\gg 1$, and $q\,r(t)$ and $q\,r(t')$ both of order $\nu$, the second term on the right oscillates very rapidly, and  hence may be neglected in the integral over $t$ and $t'$.  On the other hand, because $P(t)$ and $P(t')$ are sharply  peaked around the same time $t_L$, the argument of the first cosine on the right is small where $P(t)$ and $P(t')$ are appreciable, so this cosine may be replaced with unity.  Then (now dropping the distinction between $\nu$ and $\ell$), Eqs. \eqref{eq:eeap2} and \eqref{eq:bbap2} become
\begin{eqnarray}
&&\ell(\ell+1)C^T_{EE,\ell}\rightarrow \frac{\pi^2T_0^2}{2}\int_{\ell/r_L}^\infty q^2\,dq \,\{(1+\sin^2 b_L)^2\,\cos^2 b_L\}_{\cos b_L=\ell/qr_L}\nonumber\\ &&~~~~
\times\left|\int_{r(t)>\ell/q}dt\,P(t)\,\Psi(q,t)\,\left(1-\frac{\ell^2}{q^2\,r^2(t)}\right)^{-1/4}\right|^2\;,\label{eq:eeap3}\\[.4cm]&&
\ell(\ell+1)C^T_{BB,\ell}\rightarrow \frac{\pi^2T_0^2}{2}\int_{\ell/r_L}^\infty q^2\,dq \,\{4\sin^2b_L\cos^2b_L\}_{\cos b_L=\ell/qr_L}\nonumber\\ &&~~~~
\times\left|\int_{r(t)>\ell/q}dt\,P(t)\,\Psi(q,t)\,\left(1-\frac{\ell^2}{q^2\,r^2(t)}\right)^{-1/4}\right|^2\;.\label{eq:bbap3}
\end{eqnarray}

\begin{figure}[ht]
\begin{center}
\includegraphics[height=3.7in]{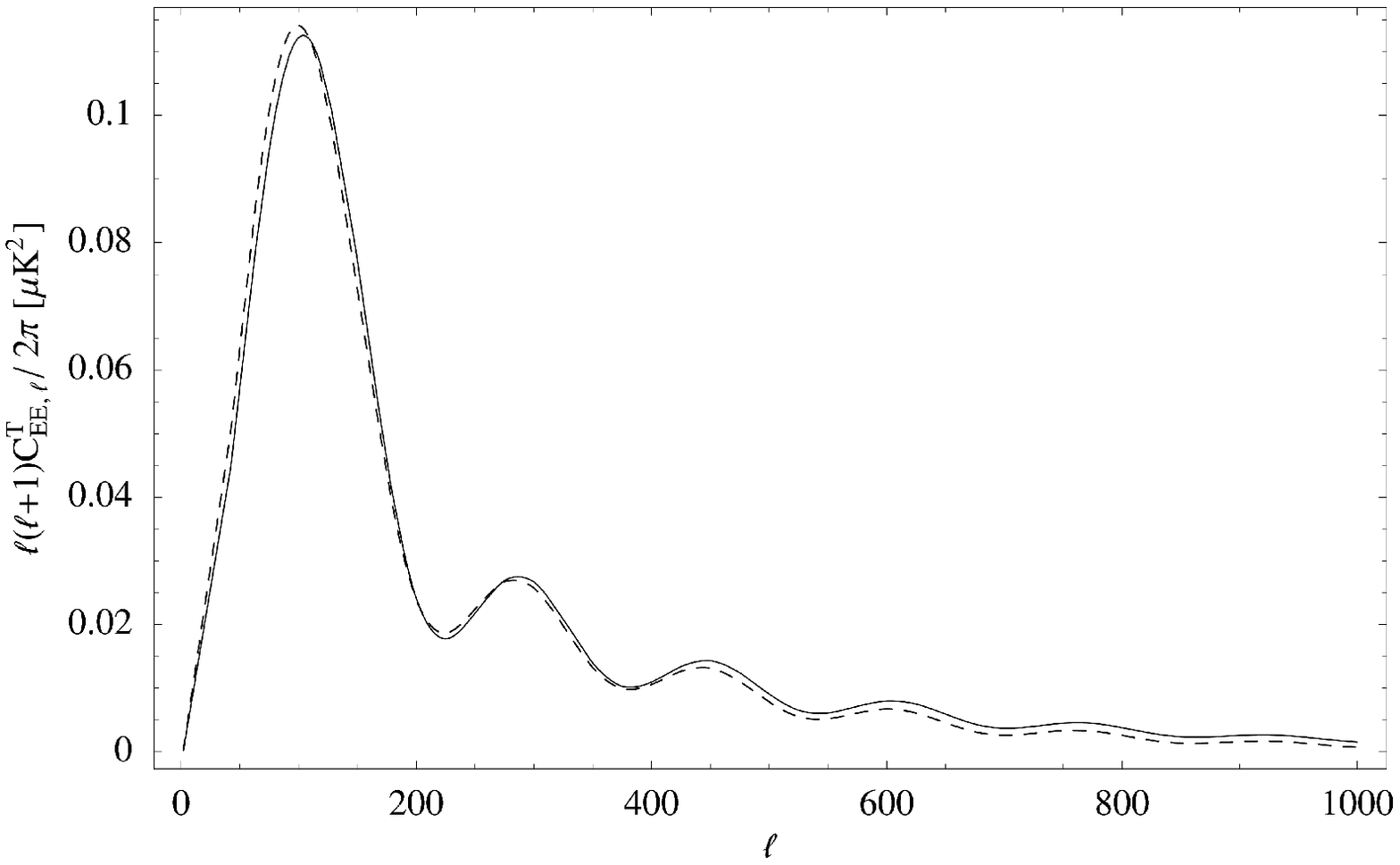}
\end{center}
\caption{Comparison of formulas for $C_{EE,\ell}^T$.  The solid line is the result of using the exact expression \eqref{eq:eeex}; the dashed line is the result of using the approximation \eqref{eq:eeap3}.  This figure shows the degree of accuracy of the further approximations used to explore the parameter-dependence of $C_{EE,\ell}^T$.}
\label{fig:eeexvsap3}
\end{figure}

\begin{figure}[ht]
\begin{center}
\includegraphics[height=3.7in]{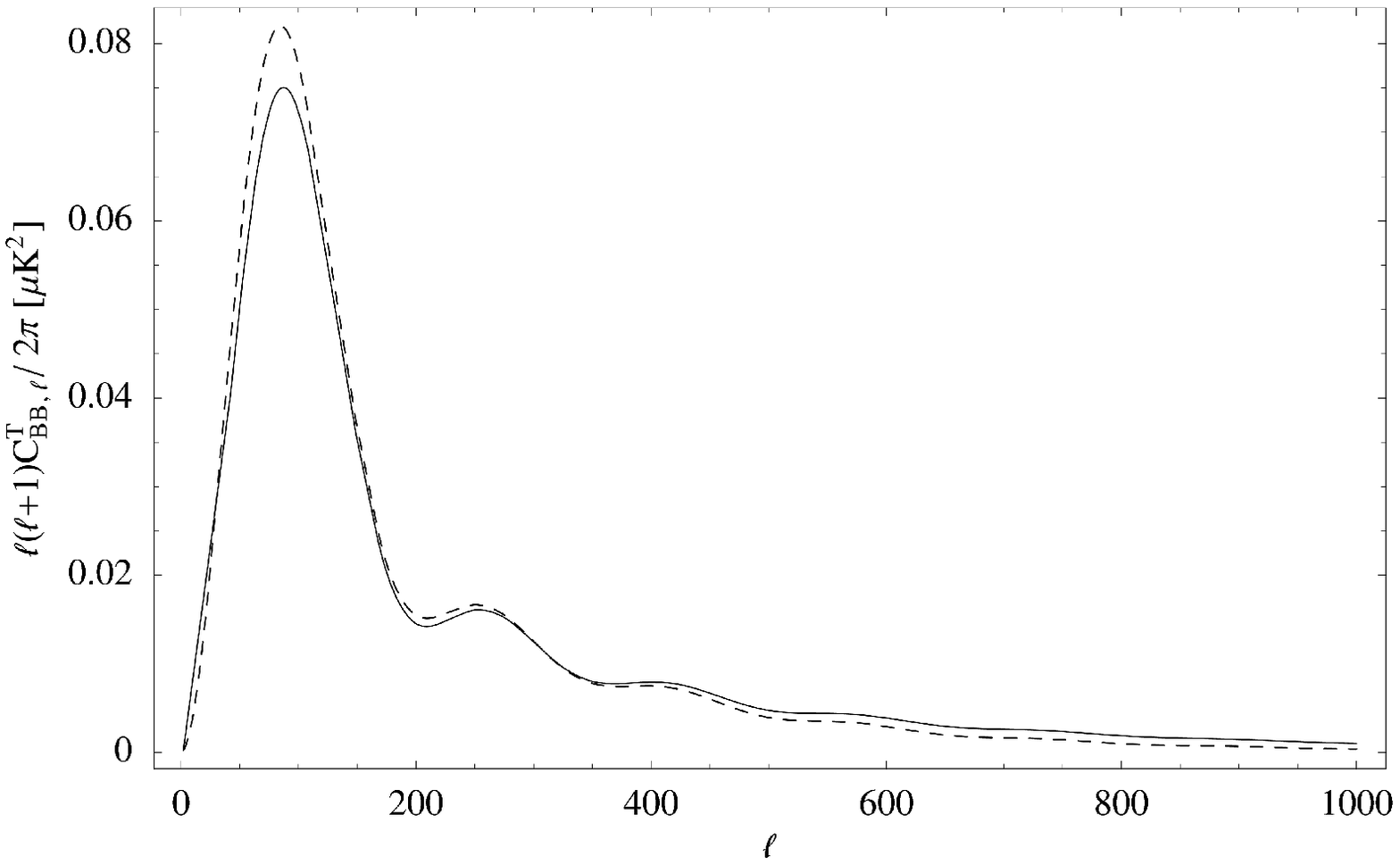}
\end{center}
\caption{Comparison of formulas for $C_{BB,\ell}^T$.  The solid line is the result of using the exact expression \eqref{eq:bbex}; the dashed line is the result of using the approximation \eqref{eq:bbap3}.  This figure  shows the degree of accuracy of the further approximations used to explore the parameter-dependence of $C_{BB,\ell}^T$.}
\label{fig:bbexvsap3}
\end{figure}

This approximation is compared with the results of the exact formulas \eqref{eq:eeex} and \eqref{eq:bbex} in Figures \ref{fig:eeexvsap3} and \ref{fig:bbexvsap3}.  As shown there, the fractional error here is less than about 20\% for $10<\ell<600$, but it becomes larger for larger values of $\ell$, where the multipole coefficients become quite small.

Eqs. \eqref{eq:eeap3} and \eqref{eq:bbap3} are useful in revealing the parameter dependence of these multipole coefficients.  
Where the last-scattering probability distribution $P(t)$ is appreciable, the only cosmological parameters on which either $P(t)$ or the source function $\Psi(q,t)$ depend  are the baryonic and matter density parameters $\Omega_Bh^2$ and $\Omega_Mh^2$, as well as the present microwave background temperature $T_0$.  All dependence of the multipole coefficients on $H_0$ or the curvature $\Omega_Kh^2$ or the vacuum energy $\Omega_\Lambda h^2$ is contained in the function $r(t)$.  But Eqs. \eqref{eq:eeap3} and \eqref{eq:bbap3} show that $r(t)$ and $ \ell$ enter in the multipole coefficients only in the combination $r(t)/\ell$.  Hence, with $\Omega_Bh^2$, $\Omega_Mh^2$, and $T_0$ fixed, to a good approximation $C^T_{EE,\ell}$ and  $C^T_{BB,\ell}$ depend on $H_0$, $\Omega_Kh^2$,  and $\Omega_\Lambda h^2$ only  through their effect on the scale of the $\ell$-dependence of $C^T_{EE,\ell}$ and  $C^T_{BB,\ell}$.  Furthermore, since $P(t)$ is sharply peaked at the time of last scattering, just as for scalar modes there is a high degree of degeneracy here: for $\ell>10$ the coefficients $C^T_{EE,\ell}$ and  $C^T_{BB,\ell}$ depend on $H_0$, $\Omega_Kh^2$,  and $\Omega_\Lambda h^2$ only through a single parameter, the radius $r(t_L)$ of the surface of last acatttering.  Of course, the degeneracy here is not as important as it is for scalar modes, because tensor modes when discovered will be studied  primarily for the purpose of measuring  the tensor/scalar ratio $r$ and the tensor slope $n_T$, rather than other cosmological parameters.

\vspace{12pt}

We are grateful to Eiichiro Komatsu for frequent helpful conversations.  This material is based upon work supported by the National Science Foundation under Grant No.  PHY-0455649 and with support from The Robert A. Welch Foundation, Grant No. F-0014.


\end{document}